\documentclass[%
 reprint,
 amsmath,amssymb,
 aps,
prl,
]{revtex4-2}
\usepackage[backref=none]{hyperref}
\usepackage{graphicx}
\usepackage{dcolumn}
\usepackage{comment}
\usepackage{lipsum}
\usepackage{xcolor}
\usepackage{physics}
\usepackage{bm}
\usepackage{soul}
\begin{document}

\preprint{APS/123-QED}

\title{Accelerated creation of NOON states with ultracold atoms via counterdiabatic driving}
\author{Simon Dengis$^{1}$}
\author{Sandro Wimberger$^{2,3}$}
\author{Peter Schlagheck$^{1}$}
\affiliation{
  $^{1}$CESAM research unit, University of Liege, B-4000 Li\`ege, Belgium}
\affiliation{
 $^{2}$Dipartimento di Scienze Matematiche, Fisiche e Informatiche,
  Universit\`a di Parma, Parco Area delle Scienze 7/A, 43124 Parma, Italy}
\affiliation{
$^{3}$INFN, Sezione di Milano Bicocca, Gruppo Collegato di Parma,
Parco Area delle Scienze 7/A, 43124 Parma, Italy
}%
\begin{abstract}
A quantum control protocol is proposed for the creation of NOON states with $N$ ultracold bosonic atoms on two modes, corresponding to the coherent superposition $\vert N,0\rangle + \vert 0,N\rangle$. This state can be prepared by using a third mode where all bosons are initially placed and which is symmetrically coupled to the two other modes. Tuning the energy of this third mode across the energy level of the other modes allows the adiabatic creation of the NOON state. While this process normally takes too much time to be of practical usefulness, due to the smallness of the involved spectral gap, it can be drastically boosted through counterdiabatic driving which allows for efficient gap engineering. We demonstrate that this process can be implemented in terms of static parameter adaptations that are experimentally feasible with ultracold quantum gases. Gain factors in the required protocol speed are obtained that increase exponentially with the number of involved atoms and thus counterbalance the exponentially slow collective tunneling process underlying this adiabatic transition.
Besides optimizing the protocol speed, our NOON-state preparation scheme achieves excellent fidelities that are competitive for practical applications.
\end{abstract}

\maketitle

The concept of entanglement is of fundamental importance in quantum physics \cite{EPR} and lies at the heart of various protocols in quantum information science \cite{NieChu}. A particularly intriguing category of entangled states are NOON states, given by a coherent superposition of the form $\vert N,0\rangle + e^{i\varphi} \vert 0,N\rangle$ that involves $N$ bosons distributed over two modes. They can be seen as microscopic Schrödinger cat states and are of great interest in the context of quantum metrology \cite{sensing1,sensing4, metrologie, sensing3, sensing2,Panda2024} especially when being realized with massive bosons.

While NOON states with up to $N \sim 10$ quanta were already produced with photons and phonons \cite{LightNOON,SCQB,Phonons}, their experimental realization with ultracold bosonic atoms is yet to be achieved. Various strategies and protocols how to prepare such NOON states with bosonic quantum gases were proposed \cite{Bec1,Bec3,Bec2,Bec4,Bec5,Bec6,Fischer2015,Bycheck2018,Pezze2019,protocolNOON}, using the interaction between the atoms of the gas as a key ingredient. Indeed, rather than being a nuisance, the presence of atom-atom interaction can be pivotally exploited to facilitate the creation of the NOON superposition since the components $\vert N,0\rangle$ and $\vert 0,N\rangle $ that form the latter can thus be energetically isolated within the system's eigenspectrum. 

\begin{figure}[!t]
    \centering
    \includegraphics[width=0.475\textwidth]{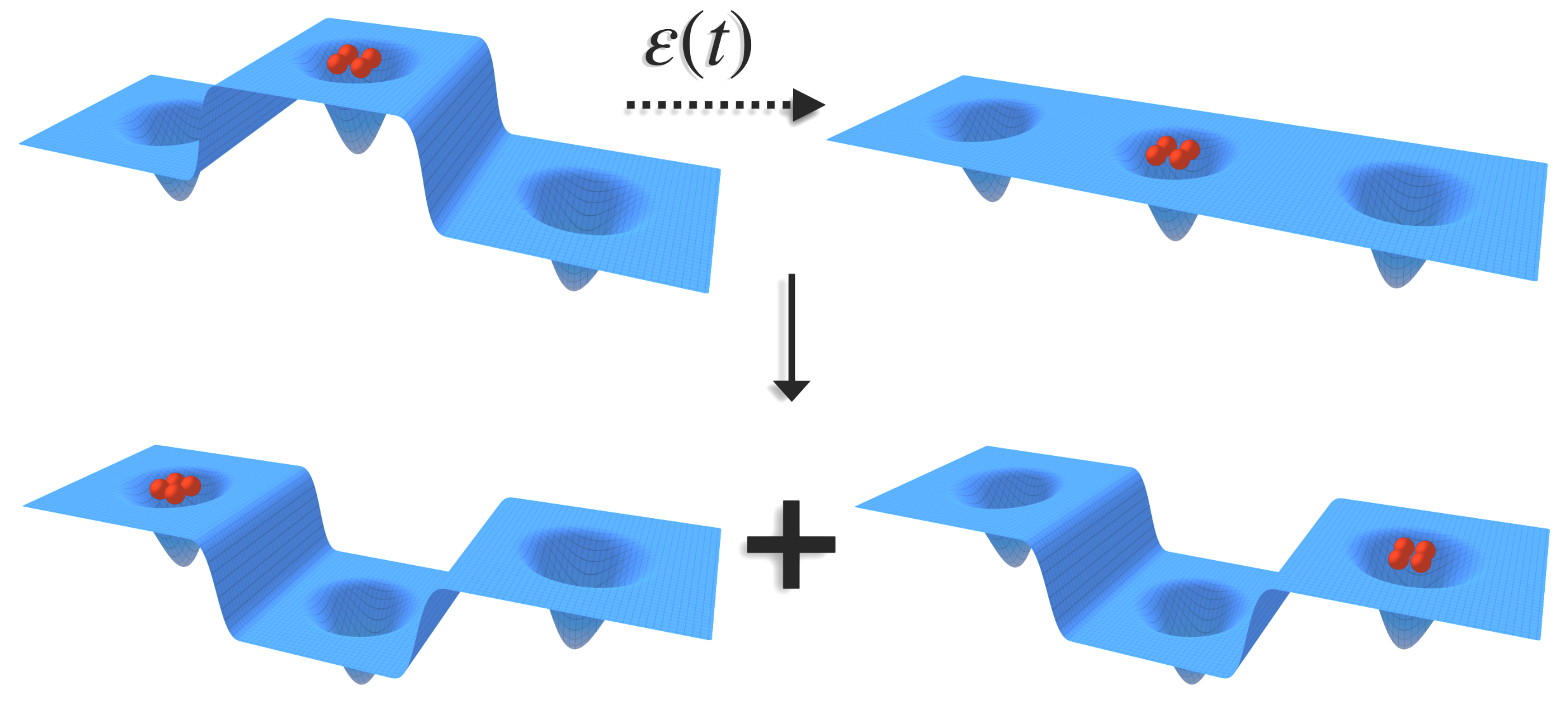}
    \caption{Schematic representation of the proposed protocol. Initially, all bosons are placed in a mode that is energetically at the top of the spectrum. Subsequently, the energy of this mode is lowered in order to achieve an avoided crossing with another energy level corresponding to the coherent superposition of the states $\vert N,0,0\rangle$ and $\vert 0,0,N\rangle$. A NOON state $\vert N,0,0\rangle + \vert 0,0,N\rangle$ is thus obtained at the end of this adiabatic transition process.}
    \label{fig:enter-label}
\end{figure}

In particular, this allows one to generate a NOON state via collective tunneling \cite{Coll1,Coll2,Coll4,Coll3}, in the parameter regime of macroscopic quantum self-trapping \cite{selftrap}, namely by preparing the system in the state $\vert N,0\rangle$ and waiting half the time that it would take to tunnel to $\vert 0,N\rangle $. Despite its appealing simplicity, this particular protocol is not suited for experimental realization since the collective tunneling time is extremely long and increases exponentially with the particle number $N$. It was recently proposed \cite{Gui1} to expose the two-mode configuration to a properly tuned periodic driving, thus giving rise to chaos-assisted tunneling \cite{chaos_assisted}. This procedure allows one to drastically enhance collective tunneling without appreciably affecting the purity of the resulting NOON superposition \cite{Gui1} and can be generalized to realize more exotic triple-NOON states on three modes \cite{Gui2}. However, it still requires a relatively long preparation time ($\sim 1$ s for ultracold $^{87}$Rb gases with $N = 5$ \cite{Gui1}) and thus imposes great technical challenges concerning the nearly perfect symmetry that must be kept between the single-particle modes as well as the shielding of the system against heating and noise effects.

To overcome the time scale problem, we employ here a different approach which is based on adiabatic transitions. To this end, we consider the presence of a third mode which is symmetrically coupled to the other modes and has a tunable single-particle energy. Initially preparing all atoms on this third mode and tuning its energy sufficiently slowly across the level of the other two modes allows one to induce an adiabatic transfer of the system's state to the desired NOON superposition forming on the other two modes. While this process also requires a long realization time, being inversely proportional to the spectral gap of the avoided level crossing, gap engineering techniques based on shortcuts to adiabaticity were developed \cite{Chen,Campo,Deffner2014,STA,delCampo2019,ieva} that can be exploited to widen the gap. We show here that an efficient counterdiabatic driving protocol can be implemented for this purpose. This protocol amounts to adaptations of the (interaction and hopping) parameters characterizing the many-body system that are static if the  time-dependent tuning of the third mode's energy is pre-optimized according to geodesic control \cite{demiplak,demirplak2,Class, geodesic}. It is experimentally feasible and allows one to generate the NOON state with high fidelity on time scales that are drastically reduced as compared to collective tunneling.

\textit{Theoretical framework}. Our system consist of $N$ bosonic atoms that are confined in a lattice with 3 wells. It is modeled by the Bose-Hubbard Hamiltonian \cite{BHModel}
\begin{align}\label{BHM}
    \hat{H}(t)= &\frac{U}{2}\sum_{i=1}^{3}\hat{a}_{i}^{\dagger}\hat{a}_{i}^{\dagger}\hat{a}_{i}\hat{a}_{i} 
    -J\sum_{i=1}^{2}\left(\hat{a}^{\dagger}_{i+1}\hat{a}_{i} + \hat{a}^{\dagger}_{i}\hat{a}_{i+1}\right) \\
    \notag
    &+ \varepsilon(t)\hat{a}_{2}^{\dagger}\hat{a}_{2},
\end{align}
(with $\hat{a}_{i}^{\dagger}$ and $\hat{a}_{i}$ the creation and annihilation operators for a bosonic particle at site $i$, respectively), which is parametrized in terms of the onsite interaction strength $U$, the hopping rate $J$ between adjacent sites, and a controllable time-dependent onsite energy $\varepsilon(t)$ for the central well. Initially, this central well is populated with $N$ bosons and has positive energy, $\varepsilon(0) > 0$, i.e. we start with the state $\vert 0,N,0\rangle$ that represents the upper edge of the system's many-body spectrum (see \cite{supmat} for more detailed discussions how to prepare this state in practice).

\begin{figure}[!t]
    \centering
     \includegraphics[width = 0.475\textwidth]{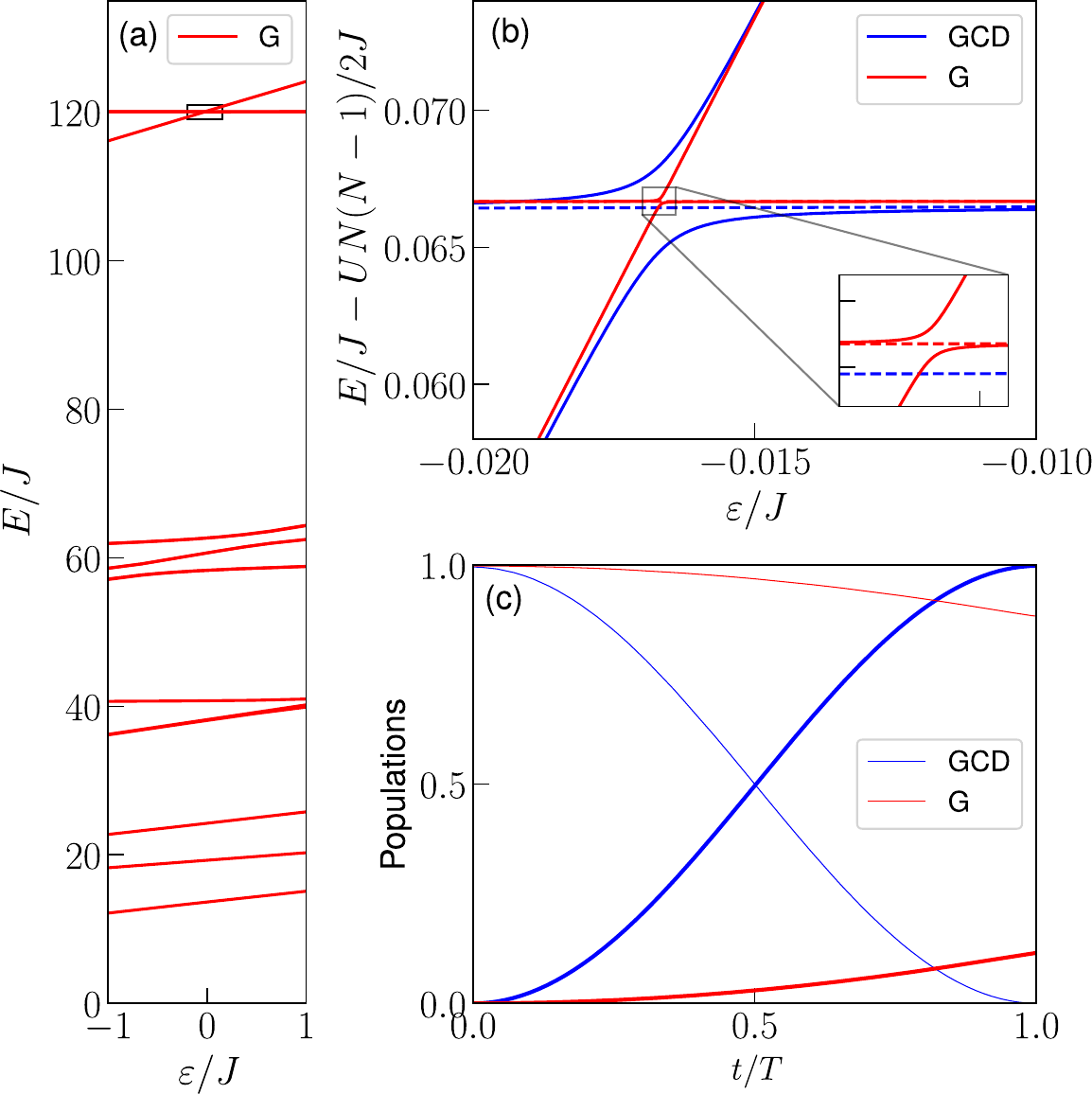}
    \caption{(a) Spectrum of the Hamiltonian (\ref{BHM}) as a function of the driving parameter $\varepsilon$, for $U = 20J$ and $N = 4$, with a geodesic driving $G$. The gap region is depicted by a black rectangle. (b) Focus on the three relevant levels of the spectrum (a) that are mainly composed of $\vert N,0,0 \rangle, \vert 0,N,0\rangle $ and $ \vert 0,0,N\rangle$. The red curves correspond to the original driven Hamiltonian (G method), while the blue curves represent the modified spectrum obtained through our GCD method for $T=3\times10^{3}\hbar/J$, which combines an optimized driving protocol with the counterdiabatic Hamiltonian. Antisymmetric states are depicted with dashed lines. (c) Detection probabilities for a fixed total protocol time $T = 3\times 10^{3}\hbar/J$. Thick curves indicates the probability of finding the system in the state $\vert \text{NOON}\rangle = (\vert N,0,0\rangle + \vert 0,0,N\rangle)/\sqrt{2}$, while thin curves relates to the probability of measuring the state $\vert0,N,0\rangle$. One can observe a nearly perfect transition with the GCD method, in contrast to method G, which only attains about 10\% of population inversion.}
    \label{fig1}
\end{figure}

This protocol is optimally executed in the self-trapping parameter regime \cite{selftrap} which is characterized by an onsite interaction strength $U$ that is large compared to the tunneling rate $J$ between the wells. In this regime, the many-body spectrum is energetically divided into multiple groups of levels. In particular, states where all particles are in the same well are all located at the top of the spectrum, isolated from the rest by a gap of size $\simeq U(N-1)/J$ (see Fig.~\ref{fig1} (a)). The presence of interaction is thus of crucial importance for our method, as it induces
a gap-protected reduced system that is useful for engineering the NOON state creation.

Due to the delocalization of states $\vert N,0,0\rangle$, $\vert 0,N,0\rangle $ and $\vert 0,0,N\rangle$, the dynamics of the system can be effectively described in terms of a 3-level system. The reduced Hamiltonian in the Fock basis $ \left(\vert N,0,0\rangle , \vert 0,N,0\rangle, \vert 0,0,N\rangle \right)$ is given by
\begin{align}\label{Hred}
    H_{\text{red}}(U,J,\varepsilon(t))= \left(
\begin{tabular}{c c c}
 $\mathcal{E}$ & $-\mathcal{J}$ & 0 \\ 
 $-\mathcal{J}$ & $\Tilde{\mathcal{E}} + \mathcal{N}\varepsilon(t)$ & $-\mathcal{J}$ \\  
 0 & $-\mathcal{J}$ & $\mathcal{E}$  \\
\end{tabular}
\right)
\end{align}
where $\mathcal{E}=\mathcal{E}(U,J)$, $\Tilde{\mathcal{E}} = \Tilde{\mathcal{E}}(U,J)$, $\mathcal{N} = N + \delta N(U,J)$ and $\mathcal{J} = \mathcal{J}(U,J)$ are obtained using perturbation theory for $NU/J \gg 1$ (see \cite{supmat}). 
The diagonal element $\Tilde{\mathcal{E}}$ differs from $\mathcal{E}$ because the central level is symmetrically coupled to the other two, while the latter only couples to the central well. However, as long as we remain within the perturbative regime where the interaction is significantly larger than the hopping, this effect is rather weak and $\Tilde{\mathcal{E}}$ will have a value very close to $\mathcal{E}$. Consequently, for $\varepsilon(t) = 0$ the three levels are very close, and the effective coupling between them is extremely small. In this regime, an adiabatic driving method can be used to steer the system towards a NOON state.

The adiabatic theorem \cite{Born} establishes that the system's state vector, when being prepared in an eigenstate of the system's Hamiltonian, will, upon sufficiently slow parameter variation, follow the parametric evolution of that eigenstate. However, this evolution will, at some point, encounter an avoided level crossing (see inset Fig.~\ref{fig1}) where the probability of a diabatic transition is high \cite{Landau,Zener,Stuck,Majorana,Berry1987,Unanyan1997,Fleischhauer1999,Lim1991}. Around this avoided crossing, it is necessary for the tuning parameter $\varepsilon$ to vary slowly, whereas away from it  the parameter variation speed can be large. 

An optimal time dependence of $\varepsilon(t)$ can be determined using the geodesics of the parameter space. To this end, we employ differential geometry to find the path towards the target state maximizing the local fidelity \cite{Provost, geodesic}. To achieve this for the $n^{\text{th}}$ excited state of the system, we require that $\varepsilon(t)$ satisfies the geodesic equation $g^{(n)}\dot{\varepsilon}(t)^{2}=\text{const.}$ \cite{demiplak, demirplak2, Class, geodesic,Kolodrubetz2017},
where
\begin{equation}
        g^{(n)} = \text{Re} \sum_{m\neq n}  \frac{\braket{n |\partial_{\varepsilon} H}{m} \braket{m | \partial_{\varepsilon} H}{n}  } {(E_{n} - E_{m})^{2}} 
\end{equation}
is the associated (single-component) metric tensor defined in terms of the instantaneous eigenstates $\vert m\rangle$ of the Hamiltonian matrix $H(t)$ and their associated eigenvalues $E_{m}$. Solving this equation provides a constraint on the driving parameter, allowing us to derive the optimal driving function :
\begin{equation}\label{geo}
    \mathcal{N}\varepsilon(t) = 2\sqrt{2}\mathcal{J}\text{tan}\left( \pi/2 - \pi t/T \right) - (\Tilde{\mathcal{E}} - \mathcal{E})
\end{equation}
for the reduced Hamiltonian (\ref{Hred}).

It is possible to enhance this process by adding a correction term to the Hamiltonian that annihilates contributions to non-adiabatic transitions \cite{Berry,Bason2012}. This term is known as the counterdiabatic Hamiltonian and can be defined as
\begin{equation}\label{hcd}
    H_{\text{CD}}(t) = i\hbar \sum_{m\neq n}\sum_{n} \frac{\langle m \vert\partial_{t}H(t)\vert n \rangle}{E_{n}-E_{m}}\vert m \rangle \langle n \vert.
\end{equation}
There exist numerous approximation proposals for the counterdiabatic Hamiltonian in complex systems \cite{Sels,Claeys, approx1,approx2}. In our case, the reduction of the problem to an effective 3-level system governed by the Hamiltonian (\ref{Hred}) allows for the exact calculation
\begin{equation}\label{redhcd}
    H_{\text{CD}}(t) = i\hbar\Omega(t) \left(
    \begin{matrix}
    0 && 1 && 0 \\
    -1 && 0 && -1 \\
    0 && 1 && 0 
    \end{matrix}\right)
\end{equation}
with $\Omega(t) = \mathcal{N}\mathcal{J}\dot{\varepsilon}(t)/[ 8\mathcal{J}^{2} + (\mathcal{N}\varepsilon(t) + \Tilde{\mathcal{E}} - \mathcal{E})^{2}] $. Note that, within that 3-level system, $\Omega$ does not depend on time if the variation of $\varepsilon$ is performed according to the above geodesic optimisation, due to the fact that the diagonal elements of $H_{\text{CD}}^{2}$, following Eq.~(\ref{hcd}), fulfill the relation $\langle n | H_{\text{CD}}^{2} | n\rangle = -\hbar^{2}g^{(n)}\dot{\varepsilon}(t)^{2}$ and are thus constant owing to the geodesic condition $g^{(n)}\dot{\varepsilon}(t)^{2}=\text{const}$. By inserting the expression~(\ref{geo}) for the geodesic driving $\varepsilon(t)$, it becomes evident that the combination of these two methods yields a time-independent counterdiabatic Hamiltonian (\ref{redhcd}) with
\begin{equation}
    \Omega = \sqrt{2}\pi/(4T).
\end{equation}

A remaining challenge is that the counterdiabatic Hamiltonian (\ref{redhcd}) is defined within the reduced 3-level system spanned by the perturbative modifications of the states $\vert N,0,0\rangle , \vert 0,N,0 \rangle$ and $\vert 0,0,N\rangle$ (see \cite{supmat}). In practice, however, it needs to be implemented in the framework of the Bose-Hubbard system under consideration. Similarly as in \cite{SP,SP2,Campo2}, we propose to emulate the requested effective Hamiltonian $H_{\text{red}} + H_{\text{CD}}$ via a suitable modification of the physical parameters $U, J$ and $\varepsilon$ of our system, thus incorporating the action of the counterdiabatic Hamiltonian into experimentally controllable parameters. We therefore define effective parameters $U_{\text{eff}}, J_{\text{eff}} $ and $\varepsilon_{\text{eff}}(t)$ such that 
\begin{equation}
    H_{\text{red}}(U,J,\varepsilon(t)) + H_{\text{CD}}  \stackrel{!}{=}H_{\text{red}}\left(U_{\text{eff}}, J_{\text{eff}},\varepsilon_{\text{eff}}(t)\right). 
\end{equation}
Applying perturbation theory (see \cite{supmat}), this identification leads to a set of three solvable equations that completely determine the effective parameters as a function of the particle number $N$:
\begin{align}
    &U_{\text{eff}} = U + 2\delta \mathcal{E}/[N(N-1)], \\
    &J_{\text{eff}} = U_{\text{eff}}^{(N-1)/N}[J^{N}/U^{N-1} + i\hbar\Omega(N-1)!/N]^{1/N}, \\
   &\varepsilon_{\text{eff}}(t) = \mathcal{N}\varepsilon(t)/\mathcal{N}_{\text{eff}} + [\Tilde{\mathcal{E}}(U,J) - \Tilde{\mathcal{E}}(U_{\text{eff}},J_{\text{eff}})]/\mathcal{N}_{\text{eff}},
\end{align}
with $\mathcal{N}_{\text{eff}} = \mathcal{N}(U_{\text{eff}},J_{\text{eff}})$ and where 
\begin{equation}
       \delta \mathcal{E} =  \mathcal{E}(U, J)-\mathcal{E}(U_\text{eff}, J_{\text{eff}}) 
   -N(N-1)(U-U_{\text{eff}})/2.
\end{equation}
As previously mentioned, the combination of counterdiabatic evolution with geodesic driving allows us to define $\Omega$ as time-independent. Consequently, the effective parameters $U_{\text{eff}}$ and $J_{\text{eff}}$ are also time-independent and can be properly tuned at the beginning of the protocol, and the only temporal dependence that needs to be managed is the driving of the energy of the central site. In optical lattices, this can be achieved by modulating the frequency difference between the lasers defining the lattice (see Supplemental Material \cite{supmat} and references \cite{lattice1,lattice2,lattice3}). 

\begin{figure}[!t]
    \centering
    \includegraphics[width=0.475\textwidth]{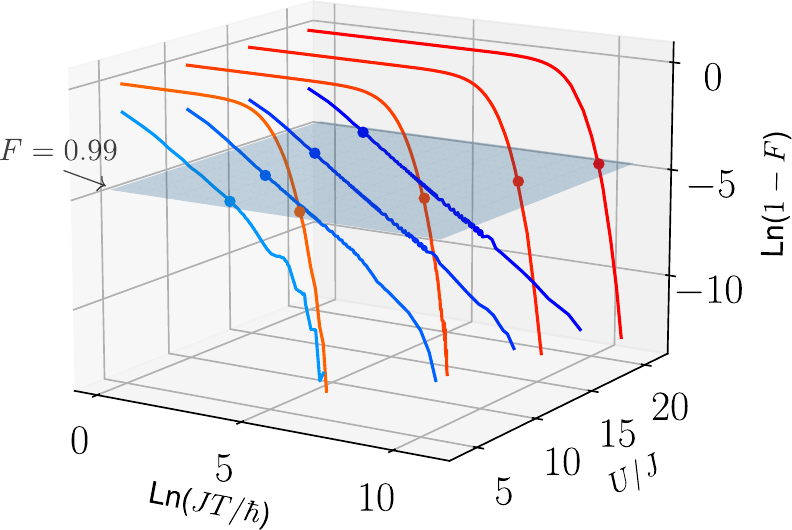}
    \caption{
    Infidelity $1-F$ as a function of the protocol time $T$ for $N=4$ and $U/J = 5,10,15$ and $20$. The fidelity $F$ is defined as the closeness of the system state to the NOON state at the end of the driving protocol. The GCD method (blue curves) achieves a NOON state for $U=20J$ with a fidelity $F>0.99$ after $T\simeq 10\hbar/J$, whereas the G method (red curves) requires a significantly longer time exceeding $T\simeq 3\times 10^{4} \hbar/J$. The plane related to a fidelity of $F =0.99$ is depicted in blue; color dots mark its intersection with the infidelity curves.
    }
    \label{dynamics_n4}
\end{figure}

\textit{Results}. 
Figure \ref{dynamics_n4} displays the numerically computed infidelity as a function of the total protocol time $JT/\hbar$ for $N=4$ and different values of $U/J$. This infidelity is defined as $1-F$ with $F= \vert \langle \text{NOON}\vert\psi(T)\rangle\vert^{2}$ and $\vert \text{NOON}\rangle = (\vert N,0,0\rangle + \vert 0,0,N\rangle)/\sqrt{2}$, where $\vert\psi(t)\rangle $ is the time-dependent state vector of the system starting at $\vert \psi(0)\rangle = \vert 0,N,0\rangle$. To quantify the impact of counterdiabatic driving, we define by $T_{G}$ and $T_{GCD}$ the protocol time needed to reach the fidelity $F=0.99$ with the geodesic protocol (G) and the combined geodesic and counterdiabatic driving (GCD), respectively. For $N=4$ and $U/J = 20$, a protocol speed gain factor $g = T_{G}/T_{GCD} \simeq  10^{4}$ is calculated. Important gains in the protocol time are also obtained for high fidelities, such as $F=0.999$, as shown in Fig. \ref{dynamics_n4}.

\begin{figure}[!h]
    \centering
    \includegraphics[width=0.475\textwidth]{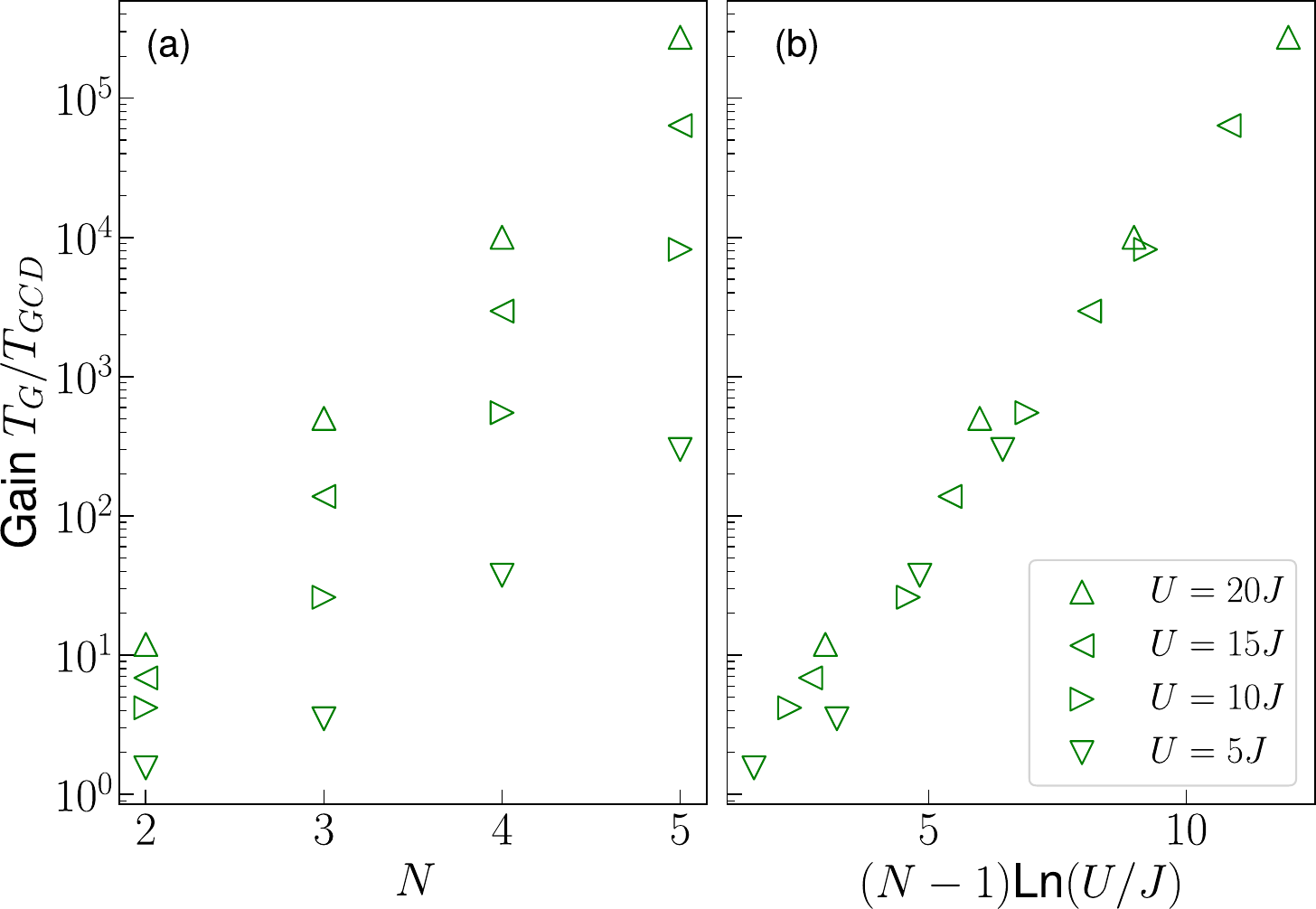}
    \vspace{10pt}
    
    \includegraphics[width = 0.475\textwidth]{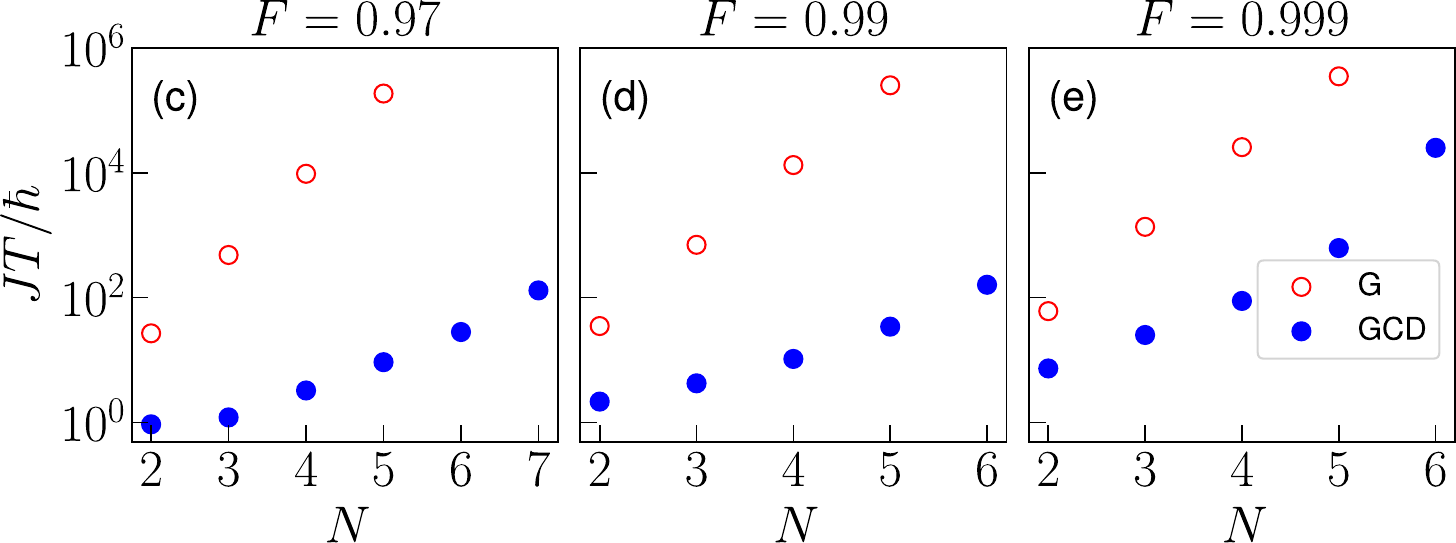}
    \caption{(a),(b) Gain ratio between the times required to achieve a fidelity of $F=0.97$ using the G and GCD methods for various interaction strengths, as a function of the number of particles. We find an exponential increase in efficiency $T_{G}/T_{GCD} \sim (U/J)^{N-1}$ when using the GCD method compared to the G method. (c),(d),(e) Protocol time $JT/\hbar$ as a function of particle number $N$ needed to yield the fidelity (a) $F=0.97$, (b) $F = 0.99$, (c) $F=0.999$ at fixed $NU/J =60$. The exponential increase of $T$ with $N$, given by $T_{G}/T_{GCD} \sim (U/J)^{N-1}$ in absence of counterdiabatic driving, is substantially amended by the GCD method.}
    \label{Gain}
\end{figure}

Figure 4 shows the time savings factor $T_{G}/T_{GCD}$ for various values of the particle number $N \leq 5$. We clearly see in Fig.~\ref{Gain}(a) that the gain increases exponentially with $N$, for various values of $U/J$. More precisely, as is seen in Fig. \ref{Gain}(b), the scaling $T_{G} \sim T_{GCD}(U/J)^{N-1}$ is obtained. This scaling reflects the inverse size of the spectral gap between the two many-body eigenstates that are formed by the $\vert 0,N,0 \rangle$ and $\vert \text{NOON}\rangle$ components in the absence of counterdiabatic driving, given by $2\sqrt{2}\mathcal{J}$ in Eq. (\ref{Hred}), which scales as $J(J/U)^{N-1}$ according to perturbation theory (see \cite{supmat}). As is shown in Fig.~\ref{Gain}(c-e), this yields an exponential increase of $T_G$ with $N$ at fixed fidelity, which is substantially amended via counterdiabatic driving.

To better understand these results in terms of a specific physical context, let us consider a gas of $^{87}$Rb atoms, with $m=1.443\times 10^{-25}$ kg and the s-wave scattering length $a_{s}=5.313$ nm, which is confined in an optical lattice that is generated by lasers with the wavelength $\lambda = 1064$ nm, whose lattice site energies are tuned to achieve the adiabatic transition (see \cite{supmat} for a more detailed description of a possible experimental realization). According to \cite{parameters2, parameters1,parameters3, Gui1}, a characteristic hopping time scale in case of the ratio $U/J = 20$ is then given by $\hbar/J = 4.4 \times 10^{-3}$ s. Using this time scale for the example case depicted in Fig.~\ref{dynamics_n4}, a fidelity of $F=0.99$ is obtained after a duration of $30$ ms with the GCD method. In contrast, using only a geodesic driving will require at least $97$ s.

\textit{Conclusion}.  We demonstrated the feasibility of generating stable NOON states using advanced adiabatic techniques. 
Specifically, the creation of NOON states can be achieved through the use of the geodesics of the parameter manifold, which is to be complemented by the addition of a time-independent counterdiabatic Hamiltonian. This latter key ingredient allows for efficient gap engineering and gives rise to a drastic reduction of the creation time as compared to a purely adiabatic transition protocol \cite{Bycheck2018}, while maintaining its inherent robustness and a highly satisfactory purity of the NOON state. In particular, for $N=4$ particles, a 99\% pure NOON state can be obtained after only a few tens of milliseconds.

The realization of NOON states with ultracold atoms in optical lattices thus becomes feasible and is facilitated by the fact that the adaptations of the interaction and hopping parameters that are needed to emulate counterdiabatic driving can be static provided that the adiabatic energy level tuning protocol is preoptimized according to geodesic control. In practice, NOON states can be produced \cite{supmat} using tunable superlattice techniques \cite{Folling}, synthetic gauge fields for inducing complex hopping \cite{Spielman,Dalibard}, quantum gas microscopes for readout \cite{Bakr,Sherson} possibly to be combined with atom conveyor belts \cite{conv2,conv1}, as well as homogeneous lattice techniques \cite{lopez} to yield the required symmetry between the states involved in the superposition. A combination with chaos-assisted collective tunneling induced by periodic driving \cite{Gui1,Gui2} and other Floquet engineering techniques \cite{Goldman2014,Bukov2015,Eckardt2017,Weitenberg2021} appears also possible and opens perspectives towards a realization with $N\sim10$ particles.
\begin{acknowledgments}
We thank N. Goldman, D. Guéry-Odelin, L. Pezzè and A. Smerzi for inspiring discussions. This project (EOS 40007526) has received funding from the FWO and F.R.S.-FNRS under the Excellence of Science (EOS) programme. S.W. acknowledges support by Q-DYNAMO (EU HORIZON-MSCA-2022-SE-01) with project No. 101131418 and by the National Recovery and Resilience Plan (PNRR), Mission 4 Component 2 Investment 1.3 -- Call for tender No. 341 of 15/03/2022 of Italian MUR funded by NextGenerationEU, with project No. PE0000023, Concession Decree No. 1564 of 11/10/2022 adopted by MUR, CUP D93C22000940001, Project title ``National Quantum Science and Technology Institute“ (NQSTI).
\end{acknowledgments}



\providecommand{\noopsort}[1]{}\providecommand{\singleletter}[1]{#1}%

\newpage

\

\newpage

\section*{Supplemental material}
\subsection*{Perturbation theory}

\setcounter{figure}{0}
\setcounter{equation}{0}



\setcounter{figure}{0}
\setcounter{equation}{0}

\begin{figure}
    \centering
    \includegraphics[width=\linewidth]{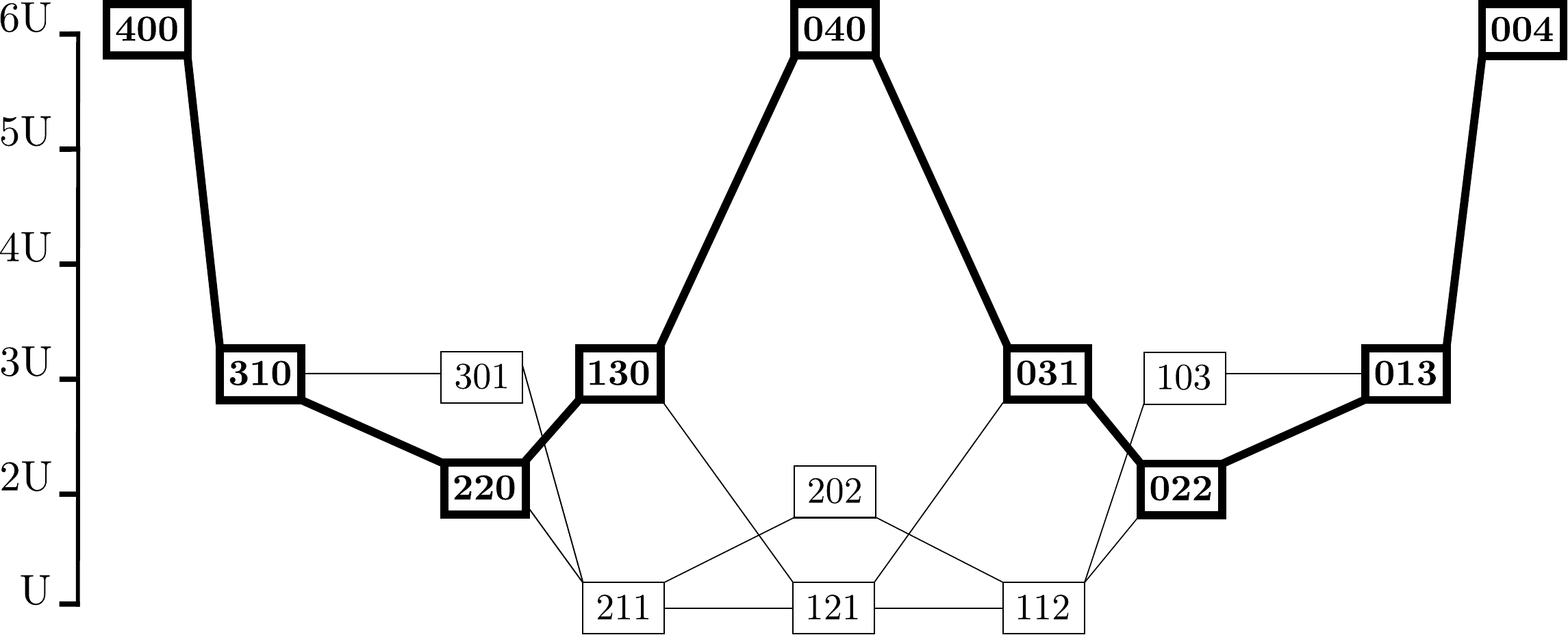}
    \caption{Schematic diagram of the various trajectories the system can take within the Hilbert space for $N=4$ particles. The vertical axis is the total interaction energy. The most probable path, i.e., the one contributing the most to the perturbative terms, is depicted in bold.}
    \label{diag}
\end{figure}

We calculate here the terms appearing in the perturbative development for $NU\gg J$.
The Hamiltonian of the full system is given by 
\begin{equation}
    \hat{H} = \frac{U}{2}\sum_{i=1}^{3}\hat{a}_{i}^{\dagger}\hat{a}_{i}^{\dagger}\hat{a}_{i}\hat{a}_{i}-J\hat{a}_{2}^{\dagger}(\hat{a}_{1} + \hat{a}_{3})-J^{*}(\hat{a}_{1}^{\dagger} + \hat{a}_{3}^{\dagger})\hat{a}_{2}
\end{equation}
for a complex hopping hopping parameter $J$.
Let us denote the interaction energy shifted by the driving term $\varepsilon$ as $E_{n_{1},n_{2},n_{3}} = n_{2}\varepsilon + \sum_{i=1}^{3} Un_{i}(n_{i}-1)/2$ and the hopping terms as $\mathcal{J}_{n_{i}, n_{2}}=J\sqrt{n_{i}(n_{2}+1)}$ and $\mathcal{J}_{n_{2},n_{i}}^{*}=J^{*}\sqrt{n_{2}(n_{i}+1)}$, where $n_{i}$ are the number of particles in site $i$.  Using the stationary Schr\"odinger equation, we end up with a system of coupled equations

\begin{equation}
    E\Psi_{N,0,0} = E_{N,0,0}\Psi_{N,0,0} - \mathcal{J}_{N, 0}\Psi_{N-1,1,0} 
\end{equation}
\begin{eqnarray}
    E\Psi_{N-1,1,0} = E_{N-1,1,0}\Psi_{N-1,1,0} - \mathcal{J}_{1,N-1}^{*}\Psi_{N,0,0} \\
    \notag - \mathcal{J}_{N-1,1}\Psi_{N-2,2,0} - \mathcal{J}_{1,0}^{*}\Psi_{N-1,0,1}
\end{eqnarray}
\begin{eqnarray}
     E\Psi_{N-2,2,0} = E_{N-2,2,0}\Psi_{N-2,2,0} - \mathcal{J}_{2,N-2}^{*}\Psi_{N-1,1,0}\\
     \notag-\mathcal{J}_{N-2, 2}\Psi_{N-3,3,0} - \mathcal{J}_{2, 0}^{*}\Psi_{N-2,1,1}
\end{eqnarray}
\begin{eqnarray}
    E\Psi_{N-1,0,1} = E_{N-1,0,1}\Psi_{N-1,0,1} - \mathcal{J}_{1, 0}\Psi_{N-1,1,0} \\
    \notag- \mathcal{J}_{N-1,0}\Psi_{N-2,1,1}
 \end{eqnarray}
 \begin{equation}
     \notag ...
 \end{equation}
\begin{eqnarray}
   E\Psi_{0,0,N} = E_{0,0,N}\Psi_{0,0,N} -\mathcal{J}_{N,0 }\Psi_{0,1,N-1},
\end{eqnarray}
for the components $\Psi_{n_{1},n_{2},n_{3}}$ that are associated with the occupation eigenstates $\vert n_{1},n_{2},n_{3}\rangle$.
This set of equations can be mapped to a set of paths that the system can take to travel through the Hilbert space from $\vert N,0,0 \rangle$ to $\vert 0,0,N\rangle$, as depicted in Fig. \ref{diag}. 

For our study, we limit ourselves to the fourth order. For this purpose, we calculate the contributions of the following paths:
\begin{align*}
    &\vert N,0,0 \rangle \leftrightarrow \vert N-1,1,0\rangle \leftrightarrow \vert N-2,2,0\rangle \\
    &\vert N,0,0 \rangle \leftrightarrow \vert N-1,1,0\rangle \leftrightarrow \vert N-1,0,1\rangle \\
    &\vert N,0,0 \rangle \leftrightarrow \vert N-1,1,0\rangle \leftrightarrow \vert N,0,0\rangle 
\end{align*}
which will add a perturbative correction $\mathcal{E}(U,J)$ to the interaction energy of the state $\vert N,0,0\rangle$ : 
\begin{equation}
    E\Psi_{N,0,0}= (E_{N,0,0} + \mathcal{E}(U,J))\Psi_{N,0,0}
\end{equation}
By solving for the considered trajectories, we obtain
\begin{eqnarray}
   \mathcal{E}(U,J)= \frac{\vert\mathcal{J}_{N,0}\vert^{2}}{E^{(2)} - \frac{\vert\mathcal{J}_{N-1,1}\vert^{2}}{E-E_{N-2,2,0}} - \frac{\vert\mathcal{J}_{1, 0}\vert^{2}}{E-E_{N-1,0,1}}}
\end{eqnarray}
where $E^{(2)} = E_{N,0,0}-E_{N-1,1,0} + \vert\mathcal{J}_{N,0}\vert^{2}/(E_{N,0,0}-E_{N-1,1,0})$. Expanding for $E \gg \vert \mathcal{J}\vert^{2}$, the corrective term is 
\begin{equation}
   \mathcal{E}(U,J) =  \frac{N \vert J \vert^{2}}{(N-1)U} + \frac{N\vert J \vert^{4}}{(N-1)^{2}(N-2)U^{3}}
\end{equation}

Since the central state $\vert 0,N,0\rangle$ is symmetrically coupled to the other two states, the value of $\Tilde{\mathcal{E}}(U,J)$ at the second order will be twice the value of $\mathcal{E}(U,J)$, but a correction appears at the fourth order. Computing this new contribution leads to identify 
\begin{align}
    \Tilde{\mathcal{E}}(U,J) = 2\mathcal{E}(U,J) - \frac{2N\vert J\vert^{4}}{(N-1)^{2}(N-2)U^{3}} \\
   \notag \times \left[\frac{N^{2} - 6N + 7}{(N-1)(2N-3)} + 1\right]
\end{align}

At this point, we are still neglecting the bias $\varepsilon$ induced by the modified energies. The full Hamiltonian is therefore $H + \varepsilon\hat{a}_{2}^{\dagger}\hat{a}_{2}$.  Indeed, $\varepsilon$ depends on time since the energy of the central well is driven according to the protocol. Consequently, the instantaneous eigenenergies of the $N$-particle system are also time-dependent. Applying a suitable gauge transformation allows us to absorb the time dependence into the central well. We then have the energies, incorporating corrective terms and the driving:  
\begin{eqnarray}
    \Tilde{\Sigma}(U,J,\varepsilon) = UN(N-1)/2 + \Tilde{\mathcal{E}}(U,J) \\
    \notag  -\frac{2N\vert J \vert^{2}}{(N-1)^{2}U^{2}}\varepsilon+ \mathcal{N}\varepsilon
\end{eqnarray}
\begin{eqnarray}
    \Sigma(U,J, \varepsilon) = UN(N-1)/2 + \mathcal{E}(U,J) \\
    \notag + \frac{N\vert J \vert ^{2}}{(N-1)^{2}U^{2}}\varepsilon
\end{eqnarray}
We thus obtain the expression : 
\begin{equation}
\mathcal{N} = N+ \delta N  = N[1 - 3\vert J \vert^{2}/((N-1)U)^{2} ] \equiv \mathcal{N}(U,J).
\end{equation}
for the shifted number of particles. The effective coupling term between $|N,0,0\rangle $ and $|0,N,0\rangle $ is calculated from the trajectory depicted in bold in Fig. \ref{diag}.
By symmetry, the coupling must be the same between $\vert 0,N,0\rangle$ and $\vert 0,0,N\rangle$. We obtain
\begin{equation}
   \mathcal{J}(U,J)= -\frac{NJ^{N}}{(N-1)! U^{N-1}}
\end{equation}
With those parameters, one can fully determine the reduced Hamiltonian $H_{\text{red}}$.

\begin{figure}[!t]
    \centering
    \includegraphics[width=\linewidth]{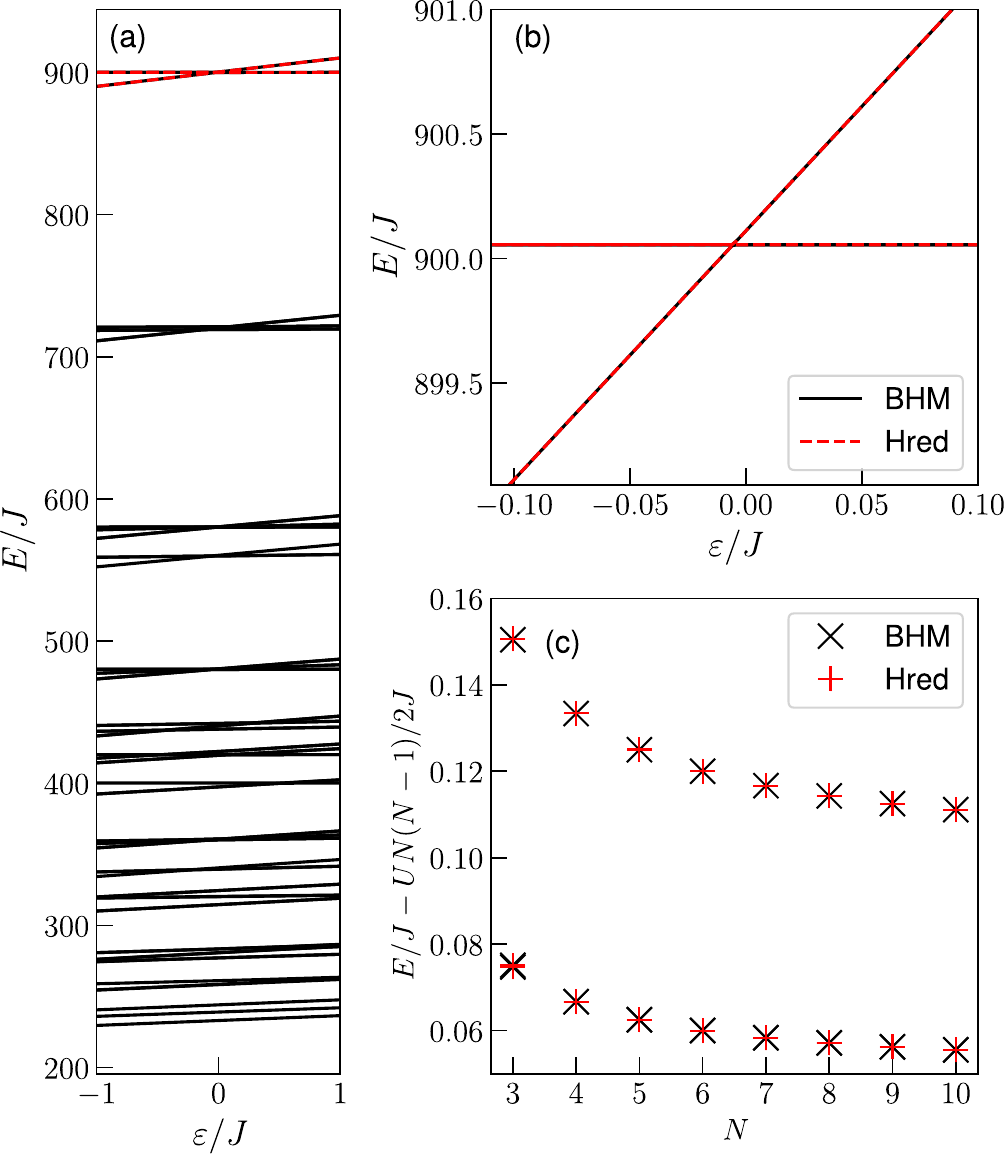}
    \caption{(a) Spectrum of the Bose-Hubbard Hamiltonian (black lines) as a function of the drive parameter $\varepsilon/J$, for $U=20J$ and $N=10$. The upper edge of the spectrum is isolated from the rest of the levels by a gap of width $U(N-1)/J$. The levels of the reduced Hamiltonian are shown with red dotted lines. (b) Zoom on the crossing of the three highest levels. (c) Value of the three highest energies shifted by the interaction energy with $U=20J$, as a function of the number of particles, for the undriven ($\varepsilon=0$) Bose-Hubbard Hamiltonian (BHM, in black) and for the reduced Hamiltonian (Hred, in red). }
    \label{fig10p}
\end{figure}

We can verify the validity of the approximation by comparing the spectrum of $H_{\text{red}}$, given by
\begin{align}
    &E_{0} = \mathcal{E}, \\
    &E_{\pm} = \left(\tilde{\mathcal{E}} + \mathcal{E} +\mathcal{N}\varepsilon \pm \sqrt{ 8 \mathcal{J}^{2} + (\Tilde{\mathcal{E}}-\mathcal{E} + \mathcal{N}\varepsilon)^{2}}\right)/2,
\end{align} 
with the three highest energy levels of the Hamiltonian (\ref{BHM}). 
Figure \ref{fig10p} (a) shows the complete spectrum of the Bose-Hubbard Hamiltonian for a particle number $N=10$. A large gap of width $U(N-1)/J$ isolates the three highest levels from the rest of the spectrum. Figure 2(b) shows a zoom on the avoided crossing between the levels of interest. The quality of the approximation (15),(16) for the energy levels is also illustrated in Fig. 2(c) where these levels are compared to the upper part of the Bose-Hubbard spectrum for various particle numbers.

\begin{figure}
    \centering
    \includegraphics[width=\linewidth]{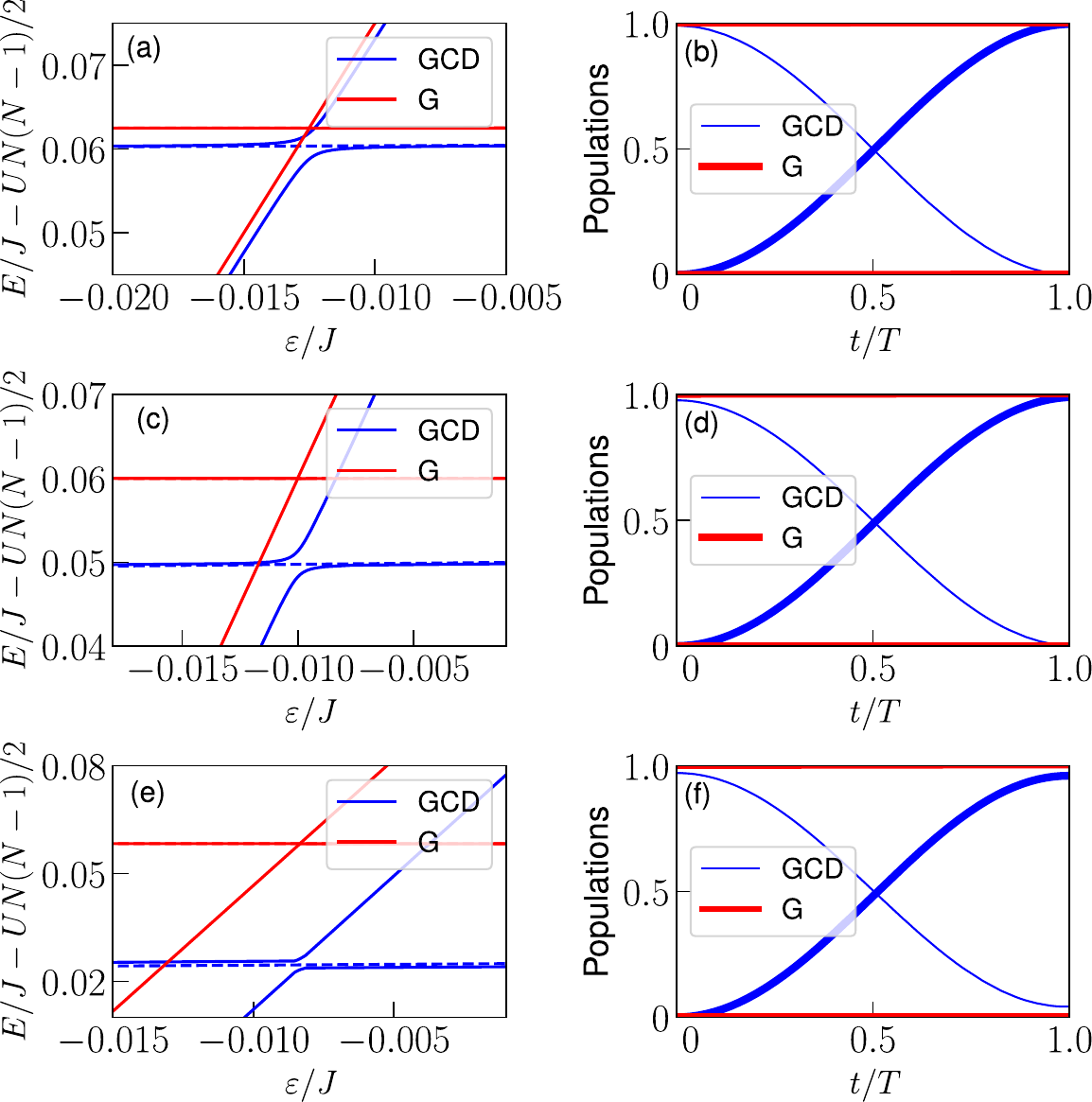}
    \caption{Left column: avoided crossing for (a) $N=5$, (c) $N=6$, and (e) $N=7$, in the case of a geodesic driving (red curve, G) and for the complete geodesic and counterdiabatic protocol (blue curve, GCD) defined for $T=10^{3}\hbar/J$.
    Right column: populations of the states $\vert 0,N,0 \rangle$ (thin lines) and $\vert \text{NOON}\rangle = (\vert N,0,0 \rangle + \vert 0,0,N \rangle)/\sqrt{2}$ (thick lines) as a function of time for (a) $N=5$, (c) $N=6$, and (e) $N=7$, for (a and b) $T=10^{4}\hbar/J$, and (c) $T = 5\times 10^{4}\hbar/J$, calculated in the framework of the complete Bose-Hubbard model. On these time scales, a nearly perfect transition between the states $\vert 0,N,0\rangle$ and $\vert \text{NOON}\rangle$ occurs for the GCD protcol (blue lines), while a vanishing population of the $\vert \text{NOON}\rangle$ state is found with the geodesic protocol (red lines).}
    \label{fign}
\end{figure}
Fig.~2 shows the impact of the counterdiabatic term on the spectrum of the Bose-Hubbard Hamiltonian. In particular, the use of a GCD protocol indeed allows for widening the gap, proportional to $J(J/U)^{N-1}$, and thus reduces the time required for adiabatic dynamics. One can also observe that the counterdiabatic term shifts the energies, since the spectrum is modified by the addition of the term ensuring adiabaticity. This effect is easily verified by examining the difference in value between the antisymmetric level of the G protocol, equal to $\mathcal{E}(U,J)$, and that of the GCD protocol, equal to $\mathcal{E}(U_{\text{eff}},J_{\text{eff}})$. The right column shows that the fidelity obtained with the GCD protocol is excellent.

\subsection{Specific proposal for a realization of a NOON state lattice}

In this section, we outline a specific quantitative proposal how to realize a periodic lattice of NOON states with bosonic atoms.

We consider for this purpose a gas of ultracold bosonic atoms that is prepared within a three-dimensional optical square lattice.
The system is supposed to be in the Mott insulator regime, where inter-site hopping is totally suppressed along two of the three lattice axes and also rather weak compared to on-site atom-atom interaction for the remaining third axis.
Along this third axis, a superlattice configuration involving three periods is to be implemented, quantitatively described, e.g., by the effective potential
\begin{equation}
  V(x) = V_0 \left[ 0.5 \cos(k x) - 2 \cos(2 k x) + \eta(t) \cos( k x / 2) \right] \label{eq:V}
\end{equation}
where $V_0$ is a suitable lattice height scale, $k$ is the characteristic lattice wave number, and $\eta(t)$ describes the dimensionless amplitude of the lattice component with the largest period $4\pi / k$.
Tuning $\eta$ as a function of time in a precisely controlled manner allows one then to induce the adiabatic transition through which NOON states can be created.
\begin{figure}[!t]
  \begin{center}
    \includegraphics[width=\linewidth]{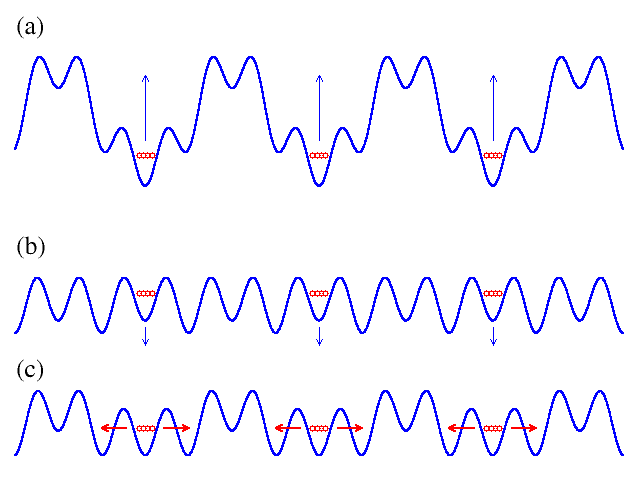}
  \end{center}
  \caption{\label{fig:lattice}
    Schematic representation of the protocol to be implemented for the creation of the NOON state. (a) The starting configuration is a superlattice with four characteristic wave numbers, featuring a distinct ground well within each superlattice cell. These ground wells are then populated with $N$ atoms per well. (b) At time $t=0$ one of the superlattice amplitudes is suddenly increased such that the former ground wells now lie energetically above their two neighbor wells. The counterdiabatic driving protocol can then be started. (c) At resonance with the neighboring wells the adiabatic transition takes place giving rise to the NOON state superposition. The blue lines in the panels represent the potential \eqref{eq:V} for the parameter (a) $\eta = -4$, (b) $\eta = 0$, and (c) $\eta = -1$.}
\end{figure}
The initial superlattice configuration ought to be such that pronounced minima appear on every fourth lattice site.
As can be seen in Fig.~\ref{fig:lattice}(a), this can be achieved e.g.\ by the choice $\eta = -4$ for the control amplitude.
This lattice is to be loaded with a gas of ultracold bosonic atoms, such that one has a mean population of $N/4$ atoms per lattice site where $N$ is the particle number for which the NOON state is to be created.
Subsequent cooling this gas to ultralow temperatures gives then rise to a Mott insulator state where every fourth site of the superlattice is populated by $N$ atoms.

At time $t=0$, a quench is to be implemented by which means the on-site energies of the populated lattice sites becomes suddenly increased, thus slightly exceeding the energies on the adjacent sites.
As illustrated in Fig.~\ref{fig:lattice}(b), this can be achieved by suddenly increasing the amplitude $\eta$ from $-4$ to $0$.
The resulting (two-period) superlattice configuration can be considered to be the starting point for the NOON state creation process to be implemented.
The amplitude $\eta$ has then to be slowly decreased, according to the protocol that we discuss in the main text of the paper.
In addition, an artificial gauge field has to be induced along the lattice \cite{DalibardSM,SpielmanSM,Eckardt20172} in order to give rise to the complex hopping matrix element (see  Eq.~(10) of the manuscript) that is required for the implementation of the counterdiabatic driving protocol.
Additional (even time-dependent) adaptations of the Bose-Hubbard parameters can, if needed, be achieved by suitable variations of the global lattice amplitude $V_0$ as well as of the analogous amplitudes along the other two axes of the lattice.

As discussed in the main part of the paper, the NOON states are then produced as soon as the populated sites of the lattice have approximately the same energy as their two neighboring sites.
As illustrated in Fig.~\ref{fig:lattice}(c), this is roughly the case for $\eta = -1$.
Further tuning the initially populated (and now depopulated) sites to lower energies then gives rise to a NOON state lattice in which every second site may or may not host $N$ atoms.
This can then be verified by quantum gas microscopes \cite{BakrSM,ShersonSM}.
A suitable atom transport protocol \cite{conv2SM,conv1SM} can then be used to direct the two components $|N,0\rangle$ and $|0,N\rangle$ of each NOON pair into spatially separate directions, for possible usage in the context of quantum metrological applications.

Note that next-to-nearest neighbor hoppings of atoms, while technically possible, can be safely neglected since they take place on time scales that are much larger than the (already very large) time scales for the creation of the NOON state. 
Maintaining nearly perfect homogeneity along the lattice, which is a key requirement for this protocol to work out correctly, is certainly a great challenge but can be, in principle, achieved using flat lattice potentials \cite{lopezSM}.


\providecommand{\noopsort}[1]{}\providecommand{\singleletter}[1]{#1}%

\end{document}